\documentstyle[12pt,epsfig]{article}
\newcommand{\la}[1]{\label{#1}}

\def\lsim{\raise0.3ex\hbox{$<$\kern-0.75em\raise-1.1ex\hbox{$\sim$}}}
\def\gsim{\raise0.3ex\hbox{$>$\kern-0.75em\raise-1.1ex\hbox{$\sim$}}}
\def\half{{\textstyle{\frac 12}}}

\begin{document}

\begin{titlepage}
\begin{flushright}
31 March, 2001\\
JYFL-3/01\\
hep-ph/0104010
\end{flushright}
\begin{centering}
\vfill

{\bf

MULTIPLICITIES AND TRANSVERSE ENERGIES \\
IN  CENTRAL $AA$ COLLISIONS AT RHIC AND LHC\\
FROM pQCD, SATURATION AND HYDRODYNAMICS
}

\vspace{0.5cm}
 K.J. Eskola$^{\rm a,b,}$\footnote{kari.eskola,vesa.ruuskanen,kimmo.tuominen,sami.rasanen@phys.jyu.fi},
 P.V. Ruuskanen$^{\rm a,b,1}$
 S.S. R\"as\"anen$^{\rm a,1}$
 K. Tuominen$^{\rm a,1}$

\vspace{1cm}
{\em $^{\rm a}$ Department of Physics, University of Jyv\"askyl\"a,
P.O.Box 35, FIN-40351
Jyv\"askyl\"a, Finland\\}
\vspace{0.3cm}
{\em $^{\rm b}$ Helsinki Institute of Physics,
P.O.Box 64, FIN-00014 University of Helsinki, Finland\\}

\vspace{1cm} \centerline{\bf Abstract} 

We compute the particle multiplicities and transverse energies at
central and nearly central $AA$ collisions at RHIC and LHC.
The initial state is computed from
perturbative QCD supplemented by the conjecture of saturation of
produced partons.
The expansion stage is described in terms of hydrodynamics assuming
longitudinal boost invariance and azimuthal symmetry. Transverse flow
effects, a realistic list of hadrons and resonance decays are
included. Comparison with the data of the multiplicities at $\sqrt
s=56$ $A$GeV and 130 $A$GeV from RHIC is done and predictions for the
full RHIC energy and LHC energy are made for the multiplicities and
transverse energies. The reduction from the initially released minijet
transverse energy to the $E_T$ in the final state is less than in the
one-dimensional case but still dramatic: a factor of 2.7 at RHIC, and
3.6 at the LHC.
\end{centering}

\vfill
\end{titlepage}

\section{Introduction}

The idea of parton saturation, the growth of the number of partons
becoming inhibited when a partonic system reaches a sufficient density,
was introduced some 20 years ago for $pp$ collisions \cite{GLR}.  In
the context of ultrarelativistic heavy ion collisions, the idea of
parton saturation in the wave functions of the colliding nuclei
governing and regulating the final state particle production was
discussed first in \cite{BM87}.  It was suggested that all particle
production in high energy $AA$ collisions could be computed in
perturbative QCD (pQCD) as the saturation scale $p_{\rm sat}\sim
A^{1/6}$ becomes large, $p_{\rm sat}\gg\Lambda_{\rm QCD}$.  Also a
pQCD approach supplemented by a soft QCD component \cite{KLL,EKL89}
implied that the perturbative mechanism would be clearly dominant in
transverse energy production at high energies.  Particle production in
$AA$ collisions has also been modeled in terms of classical gluon
fields \cite{McLV}, and in this approach the initial state parton
saturation plays a key role as well.

Saturation of produced partons at a perturbative scale $p_{\rm sat}$
can, however, be reached even if the partons in the wave functions of
the colliding nuclei are not saturated. A consistent framework of pQCD
parton production combined with the requirement of the saturation of
the produced partons was introduced in \cite{EKRT}, referred to as EKRT
here.  Based on the pQCD cross sections, it was shown that in central
collisions the minijets with transverse momenta $p_T\ge p_{\rm sat}$,
indicating an average transverse size of $\pi/p_{\rm sat}^2$ for each
minijet, are produced into the central rapidity unit so abundantly
that the available transverse area, $\pi R_A^2$, is totally
filled. Overcrowding thus takes place and the saturation of the produced
partons occurs at the scale $p_{\rm sat}$ which grows with $\sqrt s$ and
$A$. Most importantly, at RHIC and LHC energies and for large nuclei
($A\sim 200$), the saturation happens at perturbative scales, $p_{\rm
sat}\sim 1\dots2\ {\rm GeV}\gg\Lambda_{\rm QCD}$.  
We will argue below that the average features of the produced minijet
system are similar to those of a thermal system. Based on this, we
will assume that the production time, determined by the saturation
scale as $1/p_{\rm sat}$ can also be taken as the formation time $\tau_i$ of
the (approximately) thermal QGP. With this assumption and the
saturation hypothesis the initial state of the produced QGP
can be calculated using pQCD alone. 
Especially in estimating the initial energy density,
$\epsilon=E_T^{\rm pQCD}/(\pi R_A^2 \tau_i \Delta Y)$, one can make
use of the recent progress in computation of the minijet $E_T$
production in next-to-leading order pQCD \cite{ET}.

At saturation the original minijet momentum distribution is already
lost -- as the produced gluons are interacting with each other -- and
the system has started to thermalize. It was observed in \cite{EKRT}
that the produced, saturated, minijet system, which is $\sim90$\%
gluonic, looks indeed thermal from the point of view of the number of
partons and the energy per particle.  Therefore it is plausible to
assume that the momentum distributions are close to thermal at
saturation, and hydrodynamics applies in describing the expansion
stage of the system. Assuming an entropy conserving expansion stage,
scaling laws with $\sqrt s$ and $A$ were predicted in \cite{EKRT} for
the charged particle multiplicities. These predictions agree
remarkably well with the first data from RHIC \cite{PHOBOS}, thus
justifying the use of a simple geometric saturation criterion in
central collisions. To what extent the saturation approach is
applicable in non-central collisions needs careful further analysis
and will not be discussed in this paper.

In \cite{EKRT}, predictions for the measurable final state transverse
energy were also presented. However, only longitudinal expansion in
the plasma phase, using boost-invariant scaling hydrodynamics
\cite{BJORKEN}, was considered.  Based on a comparison with a
systematic study of the effects of transverse expansion on top of the
longitudinal scaling expansion \cite{KRMcLG}, it was concluded in
\cite{EKRT} that a dramatic reduction of transverse energy due to the
energy loss through the $pdV$ work in the fast initial longitudinal
expansion is expected: the estimated final $dE_T/dy$ at $y=0$ is only
$1/3$ ($1/6$) of the initially released $E_T$ of the minijets at RHIC
(LHC).  It is clear, however, that the pressure drives matter also
into the transverse direction. Therefore, we expect that the reduction
factors of $E_T$ mentioned above should be somewhat smaller when a
more realistic hydrodynamical treatment is applied.

The main goals of this paper can be stated as follows: First, we want to 
improve the estimates \cite{EKRT} of the final state particle
multiplicities by including a realistic list of hadrons and 
resonance decays in the final state. As a result, each particle will
carry on the average more than four units of entropy, and the
conversion factor between the total and charged particle multiplicity 
will be smaller than 2/3. Second, the previous estimates 
\cite{EKRT} of the measurable final state transverse energies 
are improved as we are now using a hydrodynamic description with 
transverse expansion included. 

As in \cite{EKRT}, we use the saturated minijet system as the initial
condition for the hydrodynamic expansion, including now also
transverse expansion on top of the scaling longitudinal expansion.
Our study can be applied in the central rapidity region \cite{EKR97}
but we cannot study the rapidity dependence of the observables.  We
focus on central or nearly central collisions, for which the final
state can be taken to be azimuthally symmetric. The observables in
non-central collisions, such as elliptic flow \cite{KHHH}, require a
proper treatment of an azimuthally asymmetric expansion and will be
discussed elsewhere~\cite{HKHET}.

At collider energies the minijet calculation predicts a QCD plasma in
the initial state and the hydrodynamic expansion brings the system
through mixed phase to a phase of hadron gas which finally decouples
to noninteracting hadrons. Particle spectra are computed by folding
the flow at the decoupling with the thermal motion of hadrons
according to the Cooper-Frye decoupling procedure \cite{CF}. Spectra
of all hadrons and hadron resonances up to $\Sigma(1385)$ are
calculated, followed by the resonance decays to obtain the full
spectra of stable (against strong interactions) particles. The
transverse energy and multiplicity densities, either in the pseudorapidity
$\eta$ or the real particle rapidity $y$, are calculated from the
transverse momentum spectra of final particles. Predictions for the
global observables, the number of (charged) particles and the
transverse energy are made for the energy range from the lowest RHIC
energy $\sqrt s=56$ GeV (Au-Au) to the full LHC energy $\sqrt s=5500$
GeV (Pb-Pb) with and without an effective centrality cut of 6\%
applied.  Comparison with the first data from RHIC on particle
multiplicities is shown. Also comparison with our previous results
\cite{EKRT} is made both for the multiplicities and for the transverse
energies.

\section{Initial conditions}

\subsection{Initial energy density}

As discussed in \cite{EKRT}, the average saturation scale below which the
further parton production is inhibited in a central $AA$ collision is
determined from a saturation criterion which equates the effective total
transverse area of minijets to the effective nuclear
transverse area, 
\begin{equation} N_{AA}(p_0,\sqrt s, {\bf 0}, \Delta
Y=1) \times \frac{\pi}{p_0^2} = \pi R_A^2.  \label{saturation}
\end{equation} The number of minijets produced above a transverse
momentum scale $p_0\gg \Lambda_{\rm QCD}$ into a central rapidity unit
$\Delta Y$ in an $AA$ collision with an impact parameter ${\bf
b}={\bf 0}$ and cms-energy $\sqrt s$ is computed as \begin{equation}
N_{AA}(p_0,\sqrt s, {\bf 0}, \Delta Y) = 2 T_{AA}({\bf 0})\sigma_{\rm
jet}(p_0,\sqrt s,\Delta Y,A), \label{NAA1} 
\end{equation} where
$T_{AA}({\bf 0})\approx A^2/(\pi R_A^2)$ is the standard nuclear
overlap function \cite{EKL89} and $\sigma_{\rm jet}$ is the
perturbatively computable minijet cross section with a rapidity
acceptance $\Delta Y$. In lowest order (LO) 
\begin{equation}
\sigma_{\rm jet}(p_0,\sqrt s,\Delta Y,A) = 
K\frac{1}{2}\sum_{ijkl=\atop g,q,\bar q} \int_{p_0^2,\atop \Delta Y}dp_T^2
dy_1\,dy_2\, x_1f_{i/A}(x_1,Q^2) x_2f_{j/A}(x_2,Q^2)
\frac{d\hat\sigma}{d\hat t}^{ij\rightarrow kl}\hspace{-0.5cm},
\label{sigmajet}
\end{equation}
where the minijet rapidities are $y_{1,2}$.
The fractional momenta of the colliding
partons are $x_{1,2}$, and the scale is chosen as $Q=p_T$, the
transverse momentum of produced minijets.  The
parton distributions $f_{i/A}(x,Q^2)=R_i^A(x,Q^2)f_i(x,Q^2)$ contain
nuclear effects (shadowing) in $R_i^A(x,Q^2)$ as given by the
DGLAP analysis EKS98 \cite{EKS98}.  More details of the treatment of
various subprocesses can be found e.g. in \cite{EK96}.  The
next-to-leading order (NLO) contributions to the number of
minijets are simulated by a $K$-factor taken from an exact
NLO calculation of the minijet transverse energy in Ref.~\cite{ET}
at the full RHIC and LHC energies.  As the parton distributions
in the free proton, we use the GRV94 parton distributions
\cite{GRV94}.  It should be noticed that the magnitude of the
$K$-factor depends on $\sqrt s$ and on the parton distributions
used \cite{ET}.

The saturation scale is obtained as the solution $p_{\rm sat}=p_0
(\sqrt s,A)$ of Eq. (\ref{saturation}).  After this we compute the
transverse energy carried by the produced quanta,
\begin{equation}
E_T^{AA}(p_{\rm sat},\sqrt s,{\bf 0},\Delta Y)= T_{AA}({\bf 0})
\sigma_{\rm jet}\langle E_T\rangle(p_{\rm sat},\sqrt s,\Delta Y,A),
\label{ET_AA}
\end{equation}
where $\sigma_{\rm{jet}}\langle E_T\rangle$ is the first $E_T$ moment of
the minijet $E_T$ distribution. For massless partons 
$E_T\equiv E\sin\theta=p_T$ and the rapidity acceptance $\Delta Y$ is
implemented by defining (here in LO) 
$E_T=p_T[\Theta(y_1\in\Delta Y)+\Theta(y_2\in\Delta Y)]$. The details
of the computation of $\sigma_{\rm{jet}}\langle E_T\rangle$ in LO can be
found in \cite{EK96} and the extension to the NLO in \cite{ET}.

The determination of $p_{\rm sat}$ also offers a possibility to determine
the average initial formation time of the QGP as
$\tau_i=1/p_{\rm sat}$. After this, we can form the average initial
energy density at $z=0$ by applying Bjorken's estimate \cite{BJORKEN},
\begin{equation}
\langle\epsilon\rangle = \frac{E_T^{AA}(p_{\rm sat},\sqrt s,{\bf 0},\Delta Y)}
{\pi R_A^2 \tau_i\Delta Y},
\end{equation}
where the rapidity $y$ of the minijet is assumed to be equal to the
space-time rapidity $y_s=\frac{1}{2}\ln [(t+z)/(t-z)]$. This implies
that the spatial width of a slice corresponding to a rapidity interval
$\Delta Y$ around $z=0$ is given as $\Delta z=\tau_i\Delta Y$.  For
hydrodynamics, however, we need the energy density with a transverse
profile, $\epsilon=\epsilon({\bf s})$. The simplest way of obtaining
this is through decomposing
$T_{AA}({\bf 0})=\int d^2s T_A(s)T_A(s)$, 
where the nuclear thickness function is obtained from the nuclear
density as $T_A(s)=\int dz\, n_A(\sqrt{s^2+z^2})$. 
We then arrive at the local energy density profile
in the transverse plane 
\begin{equation}
\epsilon({\bf s})\equiv\epsilon({\bf s},z=0) = [T_A(s)]^2 \cdot
\sigma_{\rm{jet}}\langle E_T\rangle(p_{\rm sat},\sqrt s,\Delta Y,A) \cdot 
\frac{p_{\rm sat}}{\Delta Y}.
\end{equation}
We use Woods-Saxon nuclear density profile $n_A(r)$ with central density
$n_0=0.17$~fm$^{-3}$, surface diffuseness $d=0.54$~fm and nuclear radius
$R_A=1.12A^{1/3}-0.86A^{-1/3}$. At the center, $T_A(0)\approx2R_An_0$.
Notice the azimuthal symmetry as only central collisions are considered.

\subsection{Centrality selection}

To have a realistic comparison with the experimental results, we need
to simulate the centrality selection of the experiments.  We do this
by considering central collisions of an effective nucleus, $A_{\rm
eff}<A$, determined by the centrality selection in the following way:
We write the total inelastic cross section of the $AA$ collision as
$\sigma_{\rm in}^{AA}(\sqrt s)=\int d^2b [1-\exp(-\sigma_{\rm
in}^{pp}(\sqrt s)T_{AA}({\bf b}))]$. Let $\sigma_{\rm in}^{AA}(b_r)$
be the inelastic cross section for collisions with $b\le b_r$
corresponding to a fraction $r$ ($100r=n$\%) of the total inelastic
cross section. We then have
\begin{equation}
r\sigma_{\rm in}^{AA} = \sigma_{\rm in}^{AA}(b_r)=
\int_0^{b_r}d^2b[1-\exp(-\sigma_{\rm in}^{pp}T_{AA}({\bf
b}))]\approx\pi b_r^2.
\end{equation}
The latter approximation is very good for $r\le0.1$. 
The average number of participants corresponding to the
centrality selection of a fraction $r$ can then be estimated as
\begin{equation}
\langle N_{\rm part}\rangle_{r}\approx \frac{1}{\pi b_r^2}
\int_0^{b_r^2}d^2 b N_{\rm part}(b),
\end{equation}
where the number of participants in a collision at an impact parameter
${\bf b}$ is computed from $N_{\rm part}({\bf b})=
2\int\,d^2s\,T_A(\vert {\bf b}-{\bf s}\vert) \left[1-\exp(-\sigma_{\rm
in}^{pp}(\sqrt s)T_A(s))\right]$ for symmetric collisions.

We then find $A_{\rm eff}$ by requiring the number of
participants in a central collision of two nuclei with mass number
$A_{\rm eff}$ to be equal
to $\langle N_{\rm part}\rangle_r$.
Motivated by the first PHOBOS results \cite{PHOBOS}
we apply here a centrality selection of 6\%, which leads to $A_{\rm
eff}=177\dots178$ for Au-Au collisions at RHIC, and to $A_{\rm
eff}=188\dots190$ for Pb-Pb collision at the LHC.
The information for computing the initial transverse energy profiles is
given in Table~1.

\begin{table}
\center
\begin{tabular}{|c|c|c|c|c|c|c|}
\hline
$\sqrt s$ [GeV] & $A/A_{\rm eff}$ & $p_{\rm sat}$ [GeV] & $\tau_i$ [fm] & $\sigma\langle
E_T\rangle$ [mbGeV] & $K$ & $\sigma_{\rm in}^{pp}$ [mb] \\
\hline
56  &197/177& 0.93/0.92 & 0.21/0.22 & 40.21/41.67 & 2.3 &35\\
130 &197/178& 1.08/1.06 & 0.18/0.19 & 65.17/67.38 & 2.3 &39\\
200 &197/178& 1.16/1.15 & 0.17/0.17 & 83.54/86.33 & 2.3 &42\\
500 &197/178& 1.34/1.32 & 0.15/0.15 & 131.7/136.0 & 2.0 &48\\
500 &208/188& 1.35/1.33 & 0.15/0.15 & 129.5/133.7 & 2.0 &48\\
1500&208/189& 1.58/1.56 & 0.12/0.13 & 217.0/223.5 & 1.6 &56\\
5500&208/190& 2.03/2.01 & 0.10/0.10 & 468.4/481.7 & 1.6 &67\\
\hline
\end{tabular}
\caption[1]{\protect\small The initial conditions for the cms-energies
and nuclei considered.  The effective nuclei $A_{\rm eff}$ correspond
to the 6\% centrality selection computed with the inelastic cross
sections $\sigma_{\rm in}^{pp}$ as explained in the text. Within a
column, the numbers on left correspond to central $AA$ collisions and
those on right are for central $A_{\rm eff}A_{\rm eff}$ collisions.
For the highest and lowest energies, the $K$-factors are taken to be
the same as computed in \cite{ET} for $\sqrt s=$ 200 and 5500 GeV, and
for $\sqrt s=500$ GeV in between them. These factors are already
included in the numbers for $\langle E_T\rangle$.  } \la{parameters}
\end{table}

\section{Expansion and final particle spectra}

We treat the expansion hydrodynamically. The high initial density with
high collision rate is one factor supporting the use of hydrodynamics.
However, for hydrodynamics to apply, the matter should be close to
thermal. This is the case with the initial conditions which we obtain
from the minijet calculation. The system is close to thermal
equilibrium in the following sense: Using $\tau_i=1/p_{\rm sat}$ as
the time when all the final partons have been produced, we can
calculate both the initial parton density $n_i$ and energy density
$\epsilon_i$ from the calculated number of partons and transverse
energy in unit rapidity.  For a thermal system these should yield the
same initial temperature, $T_i$. Assuming an ideal equation of state
of massless partons, we test this by solving first $T_i$ from
$\epsilon_i$ and then calculate the number density from $T_i$. The
result deviates from the initial minijet density negligibly at RHIC
energies and by few percents at the LHC energies, see the curves in
Fig.\ \ref{dNdy_fig}. For the hydrodynamic
expansion the key quantity is the ratio of pressure to energy
density. For massless particles with spherically symmetric momentum
distribution this is $1/3$ independent of the radial shape, the
dependence on $|{\bf p}|$, of the distribution. Also the main part of
the contribution to the pressure comes from momenta around the mean
value.  The tail, where the initial deviation from the thermal
distribution is probably the largest, contributes less.

We should also mention that a reasonably good description of
the experimental data on elliptic flow in terms of hydrodynamics
\cite{KHHH} gives further confidence in the hydrodynamic
approach.

We are here concerned with the expansion of central rapidity region at
collider energies. Independent of the exact shape of the rapidity
distributions they are expected to be quite flat for $\vert
y\vert\lsim 1$. The longitudinal expansion should then be well
described as scaling expansion, $v_z=z/t$, which means that the flow
rapidity equals the space-time rapidity $y_s=\half\ln[(t+z)/(t-z)]$.
Assuming azimuthal symmetry, as is the case in zero impact parameter
collisions, reduces the hydrodynamics to 1+1 dimensional problem
leaving only the radial expansion to be solved numerically
\cite{VR87}.

The extra ingredient which is needed for the hydrodynamic calculation,
is the equation of state (EOS). We use here an EOS A from Ref.\
\cite{Sollfrank_prc}, which was employed in a hydrodynamic study of
hadron spectra at SPS energy. The quark-gluon-plasma phase is treated
as an ideal gas of massless gluons and two flavours of quarks and
antiquarks. In hadronic phase hadrons and hadron resonances up to
$\Sigma(1385)$ are included and the repulsion among hadrons is
described through a mean field parameter $K=450$~~fm$^3$MeV. At the
phase transition temperature, $T_c=167$ MeV for zero net baryon
number, matter with densities between those of plasma and hadron gas
is assumed to be in an equilibrium mixed phase.

The spectra of all hadrons and hadron resonances up to $\Sigma(1385)$
are calculated when the density drops to values where the mean free
paths become similar to the dimension of the system. The decoupling
condition is defined in terms of the energy density which essentially
fixes the temperature since the values of baryon chemical potential
are small at collider energies in the central rapidity region. At the
decoupling we have $T_{\rm dec}\simeq120$ MeV. It can be expected that
the hadrons with smaller cross sections decouple at earlier time.
Similarly, the flavour changing cross sections are small and are known
to decouple earlier.  However, for the transversally integrated
quantities considered here, these effects are small and we have taken
the same decoupling condition for all particles and assumed both
kinetic and chemical equilibrium at the decoupling. For the folding of
flow and thermal motion we have employed the Cooper and Fry \cite{CF}
prescription to calculate the spectra.

To obtain the final stable (against strong interactions) hadron
spectra, two and three body decays of all hadron resonances 
\cite{next} have been included. The details of the particle
spectra, like their dependence on the mass number and energy of
colliding nuclei, will be considered in a separate
publication \cite{next}.

\section{Multiplicities and transverse energy}

From the treatment of the decoupling and resonance decays, we obtain
particle spectra $dN/d^2p_Tdy$ which are boost invariant and
azimuthally symmetric by construction. These symmetries are utilized
in the definitions of multiplicities and transverse energies below.

1. The rapidity distribution of the total number of particles, the
total multiplicity, is 
\begin{equation}
\frac{dN}{dy}\equiv \int dp_T \sum_i\frac{dN_i}{dp_T dy}\bigg|_{y=0},
\label{dNdy}
\end{equation}
where the index $i$ runs 
through all  particle species.
Due to the boost invariance $dN/dy$ is independent of $y$
and equals the rapidity distributions averaged over 
a rapidity bin $\Delta y$.

2. The pseudorapidity distribution of the total multiplicity is defined as
\begin{equation}
\frac{dN}{d\eta}\equiv\int dp_T \frac{dN}{dp_T d\eta }
= \int dp_T \sum_i J_i(\eta,p_T) \frac{dN_i(p_T,y)}{dp_T dy},
\end{equation}
where $y={\rm arsinh}(\frac{p_T}{m_{Ti}}\sinh\eta)$, and
\begin{equation}
J(\eta,p_T)=\frac{\partial y}{\partial\eta} = \frac{p}{E_i} 
= \frac{p_T}{E_{Ti}}
\end{equation}
 with the
transverse mass $m_{Ti}^2=p_T^2+m_i^2$. 
The averaged total multiplicity of particles in a pseudorapidity 
bin $\Delta\eta$ symmetric around $\eta=0$ can then be defined as
\begin{equation}
\frac{dN}{d\eta}\bigg|_{\Delta\eta}\equiv
\frac{1}{\Delta\eta}\int_{\Delta \eta}\frac{dN}{d\eta} =
\frac{2}{\Delta\eta}\int dp_T \sum_i\frac{dN_i}{dp_Tdy}\bigg|_{y=0}{\rm arsinh}(\frac{p_T}{m_{Ti}}\sinh\frac{\Delta\eta}{2}),
\label{dNdeta_ave}
\end{equation}
where in the last step the boost invariance was used.  To obtain the
charged particle multiplicity, one simply excludes all neutral
particles from the sum. When applying these results here, we will
choose $\Delta \eta$ as $|\eta|\le1$ in accordance with the PHOBOS
experiment \cite{PHOBOS},

3. We will be interested in two transverse energy quantities.  First, in
direct correspondence with the experimental definition of the $E_T$ as
the energy in a calorimeter cell at certain $\eta$ or scattering angle
$\theta$, we define $E_T=E\sin\theta$, and get the $E_T$ distribution
from the calculated particle spectra as
\begin{equation}
\frac{dE_T}{d\eta}\equiv \int dp_T \sum_i\frac{dN_i}{dp_Td\eta}E_{Ti}
= \int dp_T \sum_i\frac{dN_i}{dp_Tdy}\bigg|_{y=0} p_T.
\label{dETdeta}
\end{equation}
Note that again due to the boost invariance the result is independent 
of $\eta$, so averaging over a pseudorapidity bin $\Delta\eta$ gives 
conveniently 
\begin{equation}
\frac{dE_T}{d\eta}\bigg|_{\Delta\eta}
\equiv\frac{1}{\Delta\eta}\int_{\Delta\eta}\frac{dE_T}{d\eta} 
= \frac{dE_T}{d\eta}
\label{dETdeta_ave}
\end{equation}

The other transverse energy distribution we wish to study is defined as
\begin{equation}
\frac{dE_T}{dy}\bigg|_{y=0}\equiv \int dp_T\sum_i\frac{dN_i}{dp_T dy}\bigg|_{y=0} m_{Ti},
\label{dETdy}
\end{equation}
where we have used $E_T=m_T$ as is the case at $y=0$.

\section{Results}

\begin{figure}[hbt]
\vspace{-2cm}
\epsfysize=12cm
\centerline{\hspace{1cm}\epsffile{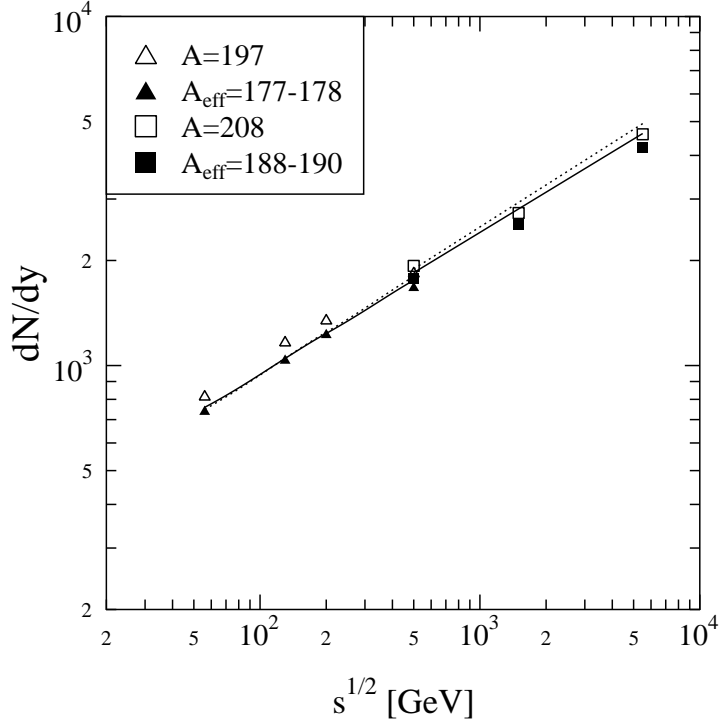}}
\caption[a]{\protect \small Total particle multiplicity $dN/dy$ for
$\sqrt s=56, 130, 500, 1500, 5500$ $A$GeV. The open circles are for
central Au-Au collisions, the open squares for central Pb-Pb
collisions.  The filled symbols are for central collisions of $A_{\rm
eff}A_{\rm eff}$ collisions, corresponding to a 6\% centrality cut
(see Table 1).  The results for central Au-Au and Pb-Pb collisions
from \cite{EKRT}, computed either from the initial multiplicity
(solid), or from the initial energy density (dotted) are shown
respectively.  }
\label{dNdy_fig}
\end{figure}

The total particle multiplicity $dN/dy$ at $y=0$, computed from
Eq. (\ref{dNdy}) by using the particle spectra, is shown in
Fig.~\ref{dNdy_fig}. We consider Au-Au collisions (triangles) at
RHIC energies $\sqrt s=56$, 130 and 200 $A$GeV and, as an interpolation
between RHIC and LHC, also 500 $A$GeV. Pb-Pb collisions (squares) are
considered at the LHC design energy $\sqrt s=5500$ (LHC) and also
at 1500 and 500 $A$GeV. The open symbols show the results for 
central collisions at ${\bf b=0}$. The filled symbols stand for the 
results with a 6\% centrality selection, described as central collisions of
effective nuclei $A_{\rm eff}<A$, see Table~1.  As seen in the figure,
the multiplicities are reduced by about 10\% by this centrality cut.

The multiplicities for central collisions from our previous work
\cite{EKRT} are shown as lines and they should be compared with the open
symbols.  In \cite{EKRT}, the final state multiplicity was obtained
either by converting the initial state parton multiplicity into
entropy, leading to $N_f(N_i)=S_i(N_i)/4=(3.6/4)*1.383A^{0.922}(\sqrt
s)^{0.383}$ (solid lines), or by converting the initial state energy density
into entropy, leading to $N_f(E_{Ti})=S_i(E_{Ti})/4=1.16A^{0.92}(\sqrt
s)^{0.40}$ (dotted lines). As observed in \cite{EKRT}, these two
results are close as the average initial energy per particle is very
near thermal.

The multiplicities obtained for central collisions in the present
study tend to be above EKRT at RHIC energies but below at the LHC
energies.  As mentioned in the introduction, this is due to different
but partly compensating effects that appear first in the conversion of
the initial energy density into the initial entropy and then the
entropy into the final state multiplicity:

\begin{itemize}

\item In \cite{EKRT}, only the gluonic degrees of freedom were used to
obtain temperature but here we choose to include also two flavours of
quarks and antiquarks. The resulting increase in the initial entropy
density (computed from $\epsilon$) enters through $s\sim
g^{1/4}\epsilon^{3/4}$, with $g$ as the effective number of the
degrees of freedom, causing a relative increase of about 23\%.
\vspace{-0.3cm}
\item In comparison with \cite{EKRT}, where no transverse profiles
were included, the total rapidity density of initial entropy,
$dS_i/dy_s$, is reduced by about 5\% due to the local conversion
$\epsilon\rightarrow s$.
\vspace{-0.3cm}
\item During the hydrodynamic evolution stage, some entropy is
generated at the phase transition and by the numerics. These effects
are quite small, e.g. at $\sqrt s=500$ $A$GeV the increase in entropy
is about 3\%.
\vspace{-0.3cm}
\item 
In a realistic hadron resonance gas of massive particles at freeze-out
(now $T_{\rm dec}=120$ MeV), the average entropy per particle is
larger than four,  $S_f/N_f\approx 4.68$ 
with the list of hadrons used. This decreases the multiplicity from 
\cite{EKRT} by about 15 \%.
\end{itemize}

Combining the effects above we get an
estimate of $dN/dy$ in terms of $N_f(E_{Ti})$ of EKRT as $dN/dy =
0.95*1.23*1.03*(4/4.68)*N_f(E_{Ti}) = 1.03*N_f(E_{Ti})$.  This
explains the small difference between the dotted lines and open
symbols in Fig.~\ref{dNdy_fig} at $\sqrt s=500$ $A$GeV, where the
$K$-factor applied coincides with the constant $K$-factor of EKRT.
Larger differences at other $\sqrt s$ result mainly from the different
$K$-factors, listed in Table~1.

\begin{figure}[hbt]
\vspace{-2cm}
\epsfysize=12cm
\centerline{\hspace{1cm}\epsffile{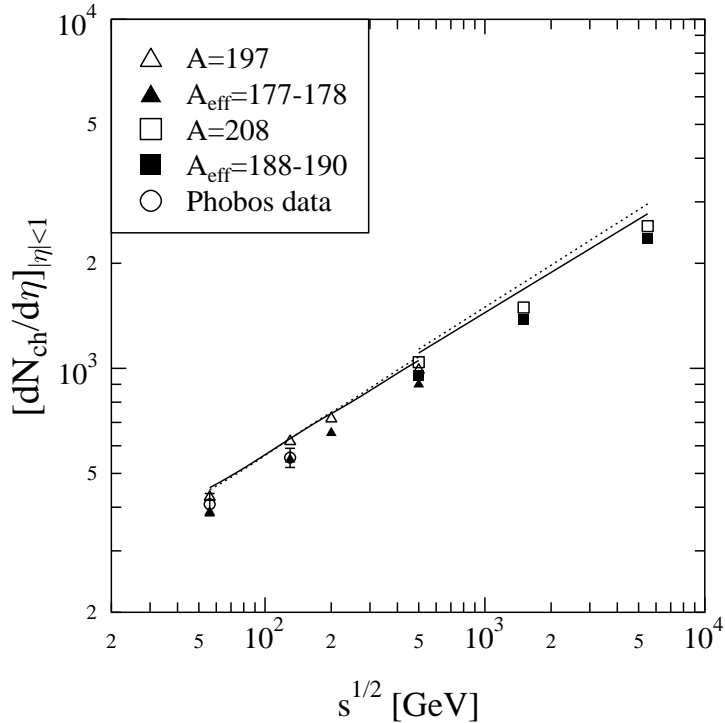}}
\caption[a]{\protect \small Charged particle multiplicity
$dN/d\eta|_{|\eta|\le1}$ averaged over a pseudorapidity bin
$-1\le\eta\le1$, computed from Eq.\ (\ref{dNdeta_ave}). The cms-energies,
the nuclei, the centrality cuts and the symbols are as in
Fig.\ \ref{dNdy_fig}.  The PHOBOS data (open circles) are shown with
systematic error bars, the small statistical error bars are not visible.  
The solid and dotted lines are again the predictions presented in \cite{EKRT} 
for the central collisions
with the same notation as in Fig.\  \ref{dNdy_fig}.
}
\label{dNdeta_fig}
\end{figure}

The corresponding charged particle multiplicities, averaged over
two central units of pseudorapidity, $dN/d\eta|_{|\eta|\le1}$,
computed from Eq. (\ref{dNdeta_ave}), are shown in Fig.~\ref{dNdeta_fig}.
Again, the open symbols denote the central collisions and the closed
ones have the 6\% centrality cut included. Otherwise the notation
is identical to that in the previous figure. The PHOBOS data at $\sqrt
s=56$ and 130 $A$GeV \cite{PHOBOS} with the 6\% centrality
selection, is shown by open circles.  The results (filled triangles) are
seen to agree remarkably well with the data, given that no fitting 
was done. 
We should mention, though, that the PHENIX data at $\sqrt s=130$ $A$GeV
(not shown in the figure) lies on the upper edge of the corresponding 
PHOBOS point. 
As indicated by the discussion above, the uncertainties 
in the theoretical computation are larger than the error bars 
of the data.

Again, the EKRT results for $dN_{\rm ch}/d\eta$ from \cite{EKRT} are
shown for the central collisions (${\bf b}={\bf 0}$) by the solid and
dotted lines, scaled down from those in Fig.~\ref{dNdy_fig} by a
factor 2/3 to account for the conversion to charged particles and a
factor 0.9 to account for the conversion from particle rapidity to
pseudorapidity and averaging over the pseudorapidity bin
$\vert\eta\vert \le 1$. The slight decrease in the ratio of the
present results (open symbols) to those of EKRT as compared with the
same ratio for total multiplicities, see Fig.~\ref{dNdy_fig}, is
mainly due to the inclusion of resonance decays, which reduces the
effective ratio $N_{\rm ch}/N_{\rm tot}$ from 2/3 to 0.6.

Note that if the EKRT results are multiplied by a further factor 0.9
to account for the 6\% centrality selection (or if $A_{\rm eff}$ is
used \cite{GW00}), they agree very well with the first RHIC data from
PHOBOS. The effects discussed above, related to the computation of the
initial entropy and conversion of the final state entropy to the final
state multiplicity, are to a good approximation constant
multiplicative factors at all energies. These factors and the varying
$K$-factors from the pQCD computation can be easily included in the
EKRT saturation model which thus provides an effective approach for
the computation of the final state multiplicities in central
collisions. If, however, particle spectra need to be computed, a more
detailed approach, such as the present one, is necessary.

\begin{figure}[hbt]
\vspace{-1.5cm}      
\epsfysize=12cm    
\centerline{\hspace{1cm}\epsffile{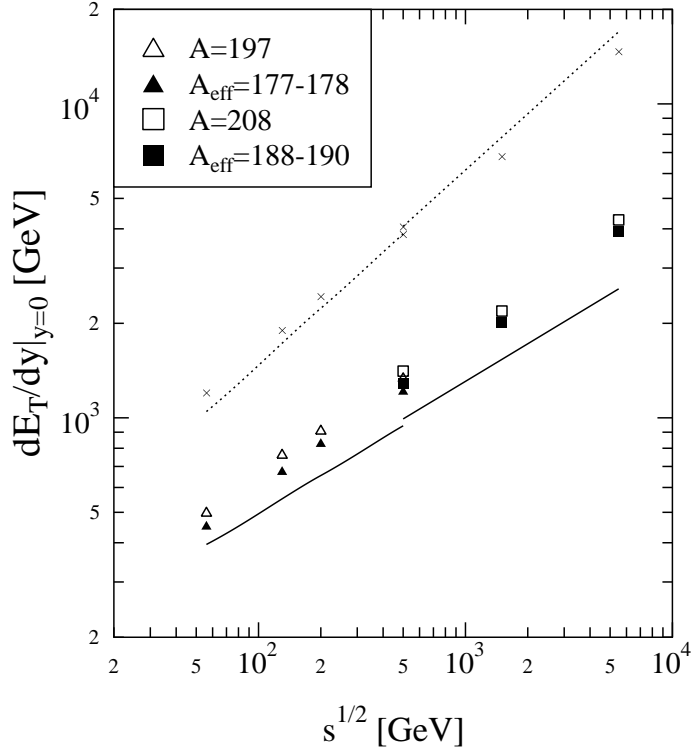}}
\vspace{-0.5cm}
\caption[a]{\protect \small The total transverse energy distribution
$dE_T/dy$ at $y=0$ computed from Eq. (\ref{dETdy}) for Au-Au and Pb-Pb
central collisions at various energies with and without a 6\%
centrality cut. The crosses show the initial $E_T^{AA}$ in $\Delta Y$
from minijets at the saturation scale.  The solid lines are the
prediction of $E_{Tf}$  from \cite{EKRT}, and the dotted lines are the
initial $E_T^{AA}$ of \cite{EKRT} for central collisions. Otherwize 
the notation is the same as in Figs.~1 and 2.  }
\label{dETdy_fig}
\end{figure}

In the computation of the final state transverse energies the
transverse expansion effects play an important role.  Next, we plot
the final state $dE_T/dy$ at $y=0$ from Eq.\ (\ref{dETdy}) 
in Fig.~\ref{dETdy_fig} for the same energies, nuclei and
centralities as in the previous figures. To get an idea of how much of
the initial transverse motion of minijets is converted into
longitudinal motion in thermalization and, especially, by the $pdV$
work in the expansion, we show $E_T^{AA}$ of minijets at
saturation by the dashed line, computed from Eq. (\ref{ET_AA}). The EKRT
results for the final state transverse energy computed in \cite{EKRT}
as $E_{Tf}=\pi R_A^2 \epsilon(\tau_f)\tau_f$ are shown by the solid
lines for central $AA$ collisions.  In \cite{EKRT} the time $\tau_f$,
at which the evolution was terminated, is the time of reaching the
energy density $\epsilon_c=\epsilon(\tau_f)$ estimated from the
results of \cite{KRMcLG}. The dashed lines show the initial $E_T^{AA}$
of \cite{EKRT} in central collisions with a constant factor $K=2$
included at all energies.  Notice how the reduction of transverse energy
between the initial and final state 
becomes smaller with the inclusion of the transverse expansion in
the hydrodynamical description. As seen in the figure, the initially
released $E_T$ is then reduced by a factor 2.7 at the full RHIC
energy, and 3.6 at the LHC.

\begin{figure}[hbt]
\vspace{-1.7cm}      
\epsfysize=12cm    
\centerline{\hspace{2cm}\epsffile{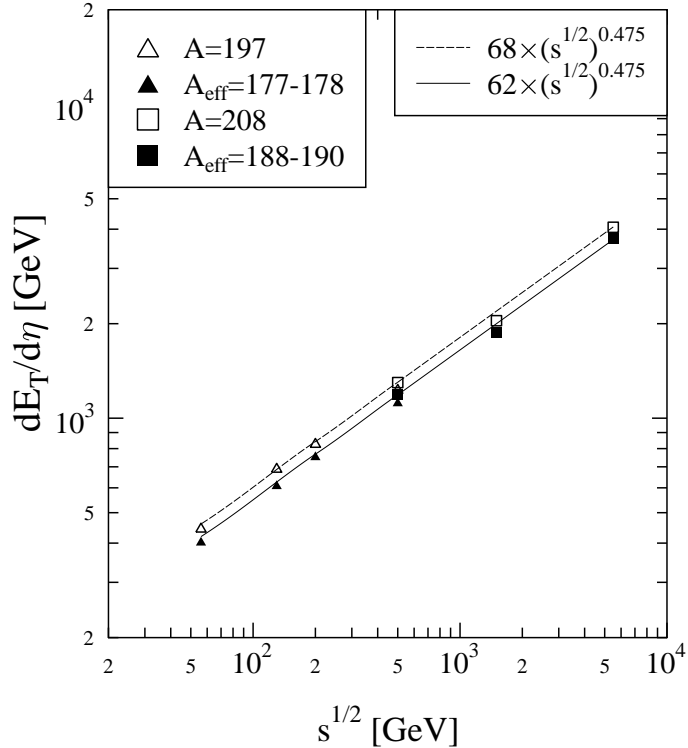}}
\vspace{-0.5cm}
\caption[a]{\protect \small Prediction for the total transverse energy
distribution $dE_T/d\eta$ at $\eta=0$ as given by Eq. (\ref{dETdeta})
for Au-Au and Pb-Pb central collisions at various energies with and
without a 6\% centrality cut. The notation is the same as in the previous
figures.  The curves are parametrizations to guide the eye. }
\label{dETdeta_fig}
\end{figure}

As the last item here, we study the measurable total transverse energy
pseudorapidity distributions of the final state hadrons.
Fig.~\ref{dETdeta_fig} shows our predictions for
$dE_T/d\eta|_{|\eta|\le1}$ averaged over the pseudorapidity bin
$|\eta|\le1$, computed from Eq. (\ref{dETdeta_ave}) for the same
energies, nuclei and centralities as before.  

We have also checked the effect of the decoupling temperature $T_{\rm
dec}$ on the multiplicities and transverse energies. If $T_{\rm dec}$
is changed by $\pm 20$ MeV, the transverse energy changes
respectively by $\pm 2\dots3$\% both at RHIC and at LHC
energies. The corresponding changes in the total multiplicities are
$\pm 1$\% both at RHIC and at LHC.  

\section{Discussion and conclusions}

We have made predictions for the measurable multiplicities and
transverse energies in central and nearly central Au-Au and
Pb-Pb collisions at collider energies. Hydrodynamics with transverse
expansion and a realistic equation of state, supplemented by
resonance decays in the final state has been applied. Contrary to the
situation at the SPS, where the measured hadron spectra were needed to
constrain the initial conditions~~\cite{Sollfrank}, we now {\em
compute} the initial conditions from perturbative QCD, supplemented
with the saturation of produced partons \cite{EKRT}. This is an essential
improvement. For the final state multiplicities the results agree very
well with the first PHOBOS results. It should be emphasized that no
fitting or fine tuning was done in order to arrive at this result.

The multiplicities obtained in this study are quite close to those of
\cite{EKRT} but the final state transverse energies are now about
20\dots40 \% larger than those predicted in \cite{EKRT}.  In comparison
with EKRT, there are different compensating effects but the difference
is mainly due to the inclusion of the transverse flow which
reduces the transverse energy loss but does not change the entropy.

It is difficult to estimate the uncertainty in the saturation approach
since the saturation criterion itself contains an undetermined
constant of order 1, containing group theory factors and powers of $\alpha_s$
and taken equal to unity in \cite{EKRT} as well as here. This uncertainty
affects the absolute normalization of the results, but we note that
data on multiplicity at one value of $\sqrt s$ and $A$ is sufficient
to fix the constant, therefore allowing  predictions for
absolute values of the global quantities at different $\sqrt s$ or
$A$. The real predictions of the saturation model are rather the
scaling laws as a function of $\sqrt s$ and $A$ for the global
quantities in central or nearly central collisions. The predicted
$\sqrt s$ scaling seems indeed to be compatible with the first PHOBOS
data \cite{PHOBOS}.

To improve these calculations further one should use the local
saturation criterion~\cite{EKT} to obtain the initial profiles. The
approach of \cite{EKT} as such, however, is not directly applicable in
this context since there one considered only the saturated part of the
initial conditions. For the hydrodynamics one needs also to take into account
the region beyond this, i.e. to consider the tails of the matter
distributions. The transition from saturated to non-saturated initial
conditions is a delicate issue concerning the interplay between
geometry and dynamics of the collision and to be studied in the future. 

As noted above, the $\sqrt s$ scaling of multiplicity in the
saturation model agrees well with the results of PHOBOS. This gives
more confidence in extrapolating the calculations towards LHC
energies. Our prediction for the average charged particle multiplicity in
central Pb-Pb collisions at $\sqrt s=5.5$ TeV is about 2560. This is slightly
smaller than the prediction of \cite{EKRT}, as the energy dependence
of the $K$-factor is taken into account.  The charged particle
multiplicity is thus clearly less than the LHC design value 8000.

While our results for the multiplicity agree with the PHOBOS data, they
are a little below the corresponding PHENIX data \cite{PHENIXN}. For
the $E_T$ our results are a little above the preliminary PHENIX data
\cite{PHENIXET}. 
In this study we have not tried to tune the calculations in order to
arrive at some desired, measured, numbers on either $N$ or
$E_T$. Rather, we have shown that the initial conditions from pQCD
minijet calculation featuring a dynamical saturation scale and
followed by hydrodynamic evolution with transverse expansion, lead 
to results which
are very similar to the measured values. More precise tests, e.g. in
terms of the transverse momentum spectra of different particles, are
needed to learn more details of the transverse dynamics and to be able
to pin down the properties of the initially produced matter.
    
\vspace{1cm}
{\it Acknowledgements:} We thank P. Huovinen, U. Heinz, P. Kolb 
and K. Kajantie for discussions and the Academy of Finland for 
financial support.


\begin{thebibliography}{9}

\bibitem{GLR}
L.~V.~Gribov, E.~M.~Levin and M.~G.~Ryskin,
Phys.\ Rept.\ {\bf 100} (1983) 1.

\bibitem{BM87}
J.~P.~Blaizot and A.~H.~Mueller,
Nucl.\ Phys.\ B {\bf 289} (1987) 847.

\bibitem{KLL}
K.~Kajantie, P.~V.~Landshoff and J.~Lindfors,
Phys.\ Rev.\ Lett.\ {\bf 59} (1987) 2527.

\bibitem{EKL89}
K.~J.~Eskola, K.~Kajantie and J.~Lindfors,
Nucl.\ Phys.\ B {\bf 323} (1989) 37.


\bibitem{McLV}
L.~McLerran and R.~Venugopalan,
Phys.\ Rev.\ D {\bf 49} (1994) 2233
[hep-ph/9309289].

\bibitem{EKRT}
K.~J.~Eskola, K.~Kajantie, P.~V.~Ruuskanen and K.~Tuominen,
Nucl.\ Phys.\ B {\bf 570} (2000) 379
[hep-ph/9909456].

\bibitem{ET}
K.~J.~Eskola and K.~Tuominen,
Phys.\ Lett.\ B {\bf 489} (2000) 329
[hep-ph/0002008];\\
``Transverse energy from minijets in ultrarelativistic nuclear 
collisions: A next-to-leading order analysis'',
hep-ph/0010319, JYFL-5-00, Phys. Rev. {\bf D} in press.

\bibitem{PHOBOS}
B.~B.~Back {\it et al.}  [PHOBOS Collaboration],
Phys.\ Rev.\ Lett.\ {\bf 85} (2000) 3100
[hep-ex/0007036].

\bibitem{BJORKEN}
J.~D.~Bjorken,
Phys.\ Rev.\ D {\bf 27} (1983) 140.

\bibitem{KRMcLG}
L.~D.~McLerran, M.~Kataja, P.~V.~Ruuskanen and H.~von Gersdorff,
Phys.\ Rev.\ D {\bf 34} (1986) 2755.

\bibitem{EKR97}
K.~J.~Eskola, K.~Kajantie and P.~V.~Ruuskanen,
Eur.\ Phys.\ J.\ C {\bf 1} (1998) 627
[nucl-th/9705015].

\bibitem{KHHH}
P.~F.~Kolb, P.~Huovinen, U.~Heinz and H.~Heiselberg,
Phys.\ Lett.\ B {\bf 500} (2001) 232
[hep-ph/0012137].

\bibitem{HKHET}
P.~F.~Kolb, U.~Heinz, P.~Huovinen, K.~J.~Eskola and K.~Tuominen,
``Centrality dependence of multiplicity, transverse energy, 
and elliptic flow from hydrodynamics'',
hep-ph/0103234.

\bibitem{CF}
F.~Cooper and G.~Frye,
Phys.\ Rev.\ D {\bf 10}, 186 (1974).

\bibitem{EKS98}
K.~J.~Eskola, V.~J.~Kolhinen and C.~A.~Salgado,
Eur.\ Phys.\ J.\ C {\bf 9} (1999) 61
[hep-ph/9807297];\\
K.~J.~Eskola, V.~J.~Kolhinen and P.~V.~Ruuskanen,
Nucl.\ Phys.\ B {\bf 535} (1998) 351
[hep-ph/9802350].

\bibitem{EK96}
K.~J.~Eskola and K.~Kajantie,
Z.\ Phys.\ C {\bf 75} (1997) 515
[nucl-th/9610015].

\bibitem{GRV94}
M.~Gluck, E.~Reya and A.~Vogt,
Z.\ Phys.\ C {\bf 67} (1995) 433.

\bibitem{VR87} 
P.~V.~Ruuskanen,
Acta Phys.\ Polon.\ B {\bf 18} (1987) 551.

\bibitem{Sollfrank_prc}
J.~Sollfrank, P.~Huovinen, M.~Kataja, P.~V.~Ruuskanen, 
M.~Prakash and R.~Venugopalan,
Phys.\ Rev.\ C {\bf 55} (1997) 392
[nucl-th/9607029].

\bibitem{next}
K.J. Eskola, H. Honkanen, P.V. Ruuskanen, S.S. R\"as\"anen and K. Tuominen,
work in progress.

\bibitem{GW00}
X.~Wang and M.~Gyulassy,
``Energy and centrality dependence of rapidity densities at RHIC'',
nucl-th/0008014.

\bibitem{Sollfrank} 
J.~Sollfrank, P.~Huovinen and P.~V.~Ruuskanen,
Eur.\ Phys.\ J.\ C {\bf 6} (1999) 525
[nucl-th/9801023].

\bibitem{EKT}
K.~J.~Eskola, K.~Kajantie and K.~Tuominen,
Phys.\ Lett.\ B {\bf 497} (2001) 39
[hep-ph/0009246].

\bibitem{PHENIXN}
K.~Adcox {\it et al.}  [PHENIX Collaboration],
``Centrality dependence of charged particle multiplicity in 
Au-Au  collisions at $\sqrt s_{\rm NN}= 130$ GeV'',
nucl-ex/0012008.

\bibitem{PHENIXET}
P.~Steinberg and W.~A.~Zajc, Talks given at ``Quark Matter 2001'', Jan
14th-20th, 2001, Stony Brook, New York;
http://www.rhic.bnl.gov/qm2001.


\end{thebibliography}
\end{document}